\documentclass[aps,prmaterials,reprint,superscriptaddress]{revtex4-2}

\usepackage{graphicx, amsmath, amssymb, xspace, braket}
\usepackage[version=4]{mhchem}
\usepackage{color}
\usepackage{lineno,hyperref}

\newcommand{\sub}[1]{\ensuremath{_{\textrm{#1}}}}

\newcommand{\Tc}{\ensuremath{T\sub{c}}\xspace}
\newcommand{\rev}[1]{#1}

\begin{document}


\title{Superconductivity in the nonsymmorphic line-nodal compound \ce{CaSb2}}


\author{Atsutoshi Ikeda}
\email[]{a.ikeda@scphys.kyoto-u.ac.jp}
\author{Mayo Kawaguchi}
\author{Shun Koibuchi}
\affiliation{Department of Physics, Kyoto University, Kyoto 606-8502, Japan}

\author{Tatsuki Hashimoto}
\affiliation{Yukawa Institute for Theoretical Physics, Kyoto University, Kyoto 606-8502, Japan}

\author{Takuto Kawakami}
\affiliation{Department of Physics, Osaka University, Toyonaka 560-0043, Japan}

\author{Shingo Yonezawa}
\affiliation{Department of Physics, Kyoto University, Kyoto 606-8502, Japan}

\author{Masatoshi Sato}
\affiliation{Yukawa Institute for Theoretical Physics, Kyoto University, Kyoto 606-8502, Japan}

\author{Yoshiteru Maeno}
\affiliation{Department of Physics, Kyoto University, Kyoto 606-8502, Japan}


\date{\today}

\begin{abstract}
We found superconductivity in \ce{CaSb2} with the transition temperature of 1.7~K 
by means of electrical-resistivity, magnetic-susceptibility, and specific-heat measurements.
This material crystallizes in a nonsymmorphic structure and is predicted to have multiple Dirac nodal lines 
in the bulk electronic band structure protected by symmetry even in the presence of spin-orbit coupling.
We discuss a possible topological superconductivity for the quasi-2-dimensional band originating mainly from one of the antimony sites.
\end{abstract}


\maketitle


\section{Introduction\label{introduction}}


%



Topological quantum phenomena have been one of the hottest topics 
in recent condensed-matter physics~\cite{Hasan2010ReviewTI, Qi2011ReviewTIandTSC, Ando2013ReviewTI}.
Many fascinating topological phenomena such as the surface Dirac cone and spin current have been actively investigated in topological insulators.
Moreover, superconductors are also classified by topology~\cite{Tanaka2012ReviewTSC, Sato2016ReviewTSC, Sato2017ReviewTSC}.
Topological superconductors are intimately related to topological insulators, and indeed topological superconductivity 
was found in a doped topological insulator~\cite{Fu2010CuxBi2Se3, Sasaki2011CuxBi2Se3, Matano2016CuxBi2Se3, Tajiri2017CuxBi2Se3}.

Nowadays, Dirac and Weyl semimetals have been attracting much attention 
as new kinds of topological materials~\cite{Yan2017ReviewWeyl, Armitage2018ReviewDiracWeyl}.
These materials are characterized by the topologically or symmetrically 
protected gapless point in the bulk electronic states in the reciprocal space.
Such gapless points lead to peculiar properties such as ultrahigh mobility, 
chiral anomaly, 
and surface Fermi arcs. 
It is also predicted that Dirac and Weyl semimetals can become 
topological superconductors~\cite{Kobayashi2015DiracSC, Li2018DiracSC, Kawakami2018TopoSC}.
Thus, superconductivity experimentally found in pressurised and doped Dirac materials~\cite{He2016Cd3As2, Oudah2016superconductivity} 
can be leading candidate for realization of topological superconductors.

More recently, the concept of the Dirac and Weyl semimetals has been extended to the line-nodal materials, in which 
the bulk energy gap closes along a line in the Brillouin zone~\cite{Fang2015NodalLine, Yamakage2016CaAgX, Kobayashi2017NodalLine}.
These materials are predicted to exhibit several notable phenomena including a long-ranged Coulomb interaction~\cite{Huh2016NodalRing}, 
a large surface polarization charge~\cite{Hirayama2017NodalLine}, and quasitopological electromagnetic response~\cite{Ramamurthy2017NodalLine}.
Furthermore, if superconductivity is induced in the materials with torus-shaped Fermi surfaces derived from nodal loops, 
topological crystalline and second-order topological superconductivity is expected to be realized~\cite{Shaourian2018NodalLoopSC}.

Although many materials can have nodal lines originating from band crossings in the absence of spin-orbit coupling, 
such nodal lines generally become gapped in the presence of spin-orbit coupling and 
the material turns into an insulator or a point-nodal semimetal~\cite{Ali2014PbTaSe2, Yamakage2016CaAgX}.
An additional symmetry such as nonsymmorphic (screw or glide) symmetry is necessary to retain the gapless nodal lines~\cite{Fang2015NodalLine}.
Thus, so far, only a limited number of materials are proposed as line-nodal materials in the presence 
of spin-orbit coupling~\cite{Carter2012NodalLine, Chen2015NodalLine, Liang2016NodalLine, Sun2017NodalLine}, 
and superconductivity in materials with such nodal lines robust against spin-orbit interaction is even more limited~\cite{Neumeier2017BaSbS3}.

In this paper, we report discovery of superconductivity with the transition temperature $\Tc=1.7$~K in the nonsymmorphic metal \ce{CaSb2}.
A band structure calculation shows that \ce{CaSb2} has several nodal lines protected by the screw and mirror symmetries 
\rev{of the space group $P2_1/m$} even in the presence of spin-orbit coupling and that some nodes cross the Fermi energy~\cite{Funada2019CaSb2}.
We present evidence of bulk superconductivity obtained by means of magnetic susceptibility, resistivity and specific heat.
The zero-temperature upper critical field is found to be around 0.2~T, corresponding to the coherence length of 40~nm.
We also discuss the possibility of topological superconductivity.
This superconductivity in a nonsymmorphic nodal-line material can trigger exploration 
of possible novel superconducting phenomena originating from the robust nodal lines.

\section{Experiment\label{experiment}}
\begin{figure*}
\includegraphics[width=0.85\linewidth]{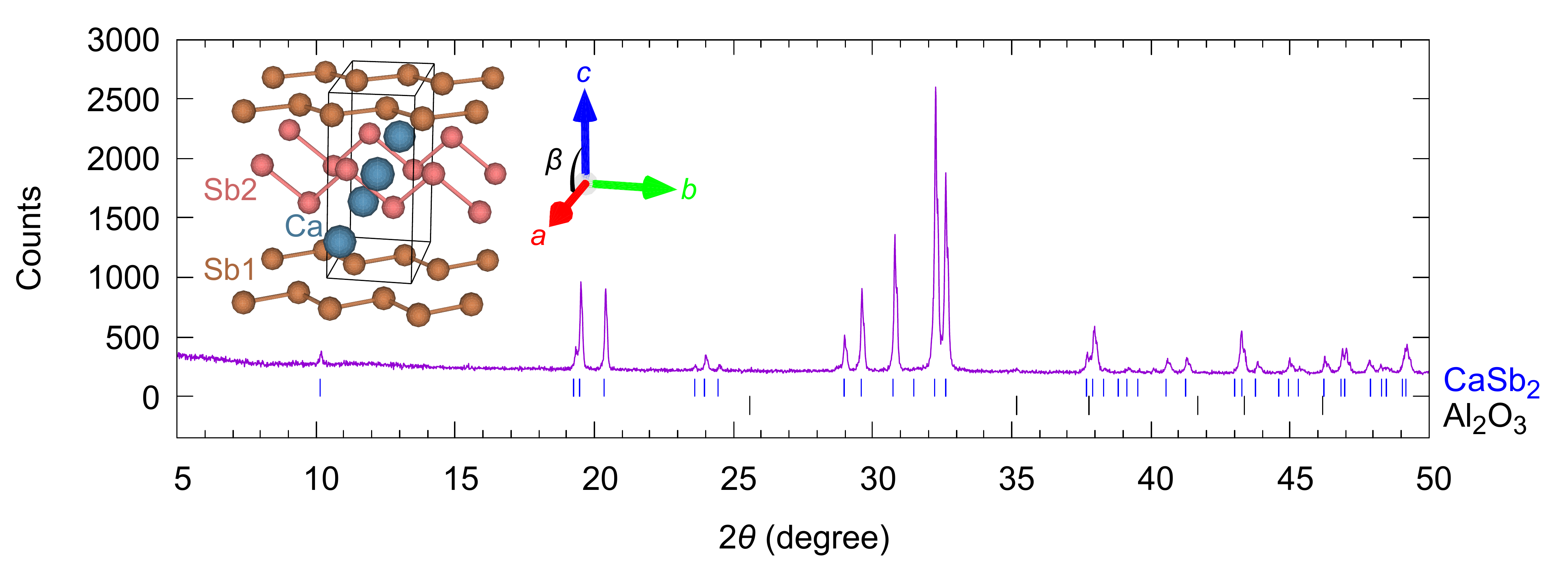}
\caption{Powder X-ray diffraction pattern of a \ce{CaSb2} sample.
The lattice parameters of \ce{CaSb2} were extracted to be $a=0.4741$~nm, $b=0.4182$~nm, 
$c=0.9073$~nm, and $\beta=106.3^\circ$, in agreement with the previous studies~\cite{Deller1976CaSb2, Funada2019CaSb2}.
The bars at the bottom indicate the reported peak positions of \ce{CaSb2} 
(blue bars; PDF 01-071-0135) and \ce{Al2O3} (black bars; PDF 01-089-7716).
For our sample, very small peaks of the latter are attributable to a contamination from crucibles.
The inset shows the crystalline structure of \ce{CaSb2} \rev{with the nonsymmorphic space group $P2_1/m$ (No.~11, $C_{2h}^2$)
with the screw axis along $b$.} 
\rev{The structural image is} produced using the program VESTA~\cite{Momma2011VESTA}.}
\label{fig: XRD}
\end{figure*}

Stoichiometric mixture of Ca (Sigma-Aldrich, 99.99\%) and Sb (Rare Metallic, 99.99\%) was placed 
in an alumina crucible (IRIE Corporation, SSA-S), and the crucible was sealed in a quartz tube under 0.2 atm of argon at room temperature.
The tube was heated up to 1,000$^\circ$C in 3 h, kept at that temperature for 24 h using a box furnace (Denken, KDF 80S), 
and quenched in water in order to prevent the phase splitting into \ce{Ca11Sb10} and Sb~\cite{Okamoto1997Ca-Sb}.
The sample was then ground and pelletized in an argon atmosphere.
The pellet was sealed again in a quartz tube under 0.3 atm of argon at room temperature.
The tube was heated to 550$^\circ$C in 2 h, kept for 48 h, and furnace cooled.
Subsequent measurements are all performed using pelletized polycrystalline samples.
This second annealing temperature was chosen to be below the \ce{CaSb2}-Sb eutectic temperature of 587$^\circ$C~\cite{Okamoto1997Ca-Sb}.
We also tried a method similar to the previously reported one, starting from a molar ratio of Ca:Sb = 1:3.9~\cite{Funada2019CaSb2} 
and slow cooling from 700$^\circ$C to 570$^\circ$C, but it resulted in a sample containing a lot more additional Sb side product.

Powder x-ray diffraction (PXRD) was measured with a commercial diffractometer equipped with an array of 192 detectors (Bruker AXS, D8 Advance)
using the Cu-$K\alpha$ radiation. 
We did not find any trace of decomposition of \ce{CaSb2} even after the sample was kept in air at room temperature for more than two weeks.

Direct-current (DC) magnetization was measured with a commercial magnetometer 
using a superconducting quantum interference device (Quantum Design, MPMS-XL) 
equipped with a \ce{^3He} refrigerator (IQUANTUM, iHelium3).
\rev{A cut pellet of 50.6~mg (approximately $3.3\times3.4\times1.0$~mm$^3$) was used.}
Alternating-current (AC) magnetic susceptibility was measured with a lock-in amplifier 
(Stanford Research Systems, SR830) using a miniature susceptometer~\cite{Yonezawa2015ADR} 
compatible with an adiabatic demagnetization refrigerator (ADR) of a commercial apparatus (Quantum Design, PPMS).

The electrical resistivity was measured by a standard four-probe method using the ADR option of PPMS.
Gold wires with a diameter of 25~$\mu$m 
were attached to the pellet using room-temperature-cure silver paste (DuPont Electronics Material, 4929N) 
with $n$-butyl acetate 
as solvent.
We comment that zero resistance was not observed when we used silver epoxy (Epoxy Technology, H20E), possibly because 
the sample reacted with air or with the silver epoxy during the curing process at 120$^\circ$C\@.

Heat capacity \rev{from 0.42 to 3.4~K} was measured using a calorimeter 
for a \ce{^3He} option of PPMS, by using the relaxation method.
A cut pellet of 15.7~mg (approximately $2\times2\times1$~mm$^3$) was fixed 
on the sample stage with a thin layer of grease (Apiezon, N-grease) to obtain good thermal connection.
For each measurement, a heat pulse raised the sample temperature by 3\% of the bath temperature and 
the temperature relaxation curve was fitted using a double exponential function to evaluate the heat capacity.

\section{Results\label{results}}
First, we present the PXRD pattern in Fig.~\ref{fig: XRD}.
The pattern revealed that our \ce{CaSb2} sample is single phased, 
except for a slight \ce{Al2O3} impurity attributable to contamination from the crucible.
The lattice parameters of the \ce{CaSb2} phase were extracted to be $a=0.4741$~nm, $b=0.4182$~nm, $c=0.9073$~nm, 
and $\beta=106.3^\circ$ \rev{with the space group $P2_1/m$ (No.~11, $C_{2h}^2$)} by the Rietveld analysis with a weighted pattern $R$ value $R\sub{wp}=12.7$\% and goodness of fit $S=1.21$.
The fitting errors were about 0.002\%.
These lattice parameters are consistent with the previous reports within 0.06\%~\cite{Deller1976CaSb2, Funada2019CaSb2}.

\begin{figure}
\includegraphics[width=0.85\linewidth]{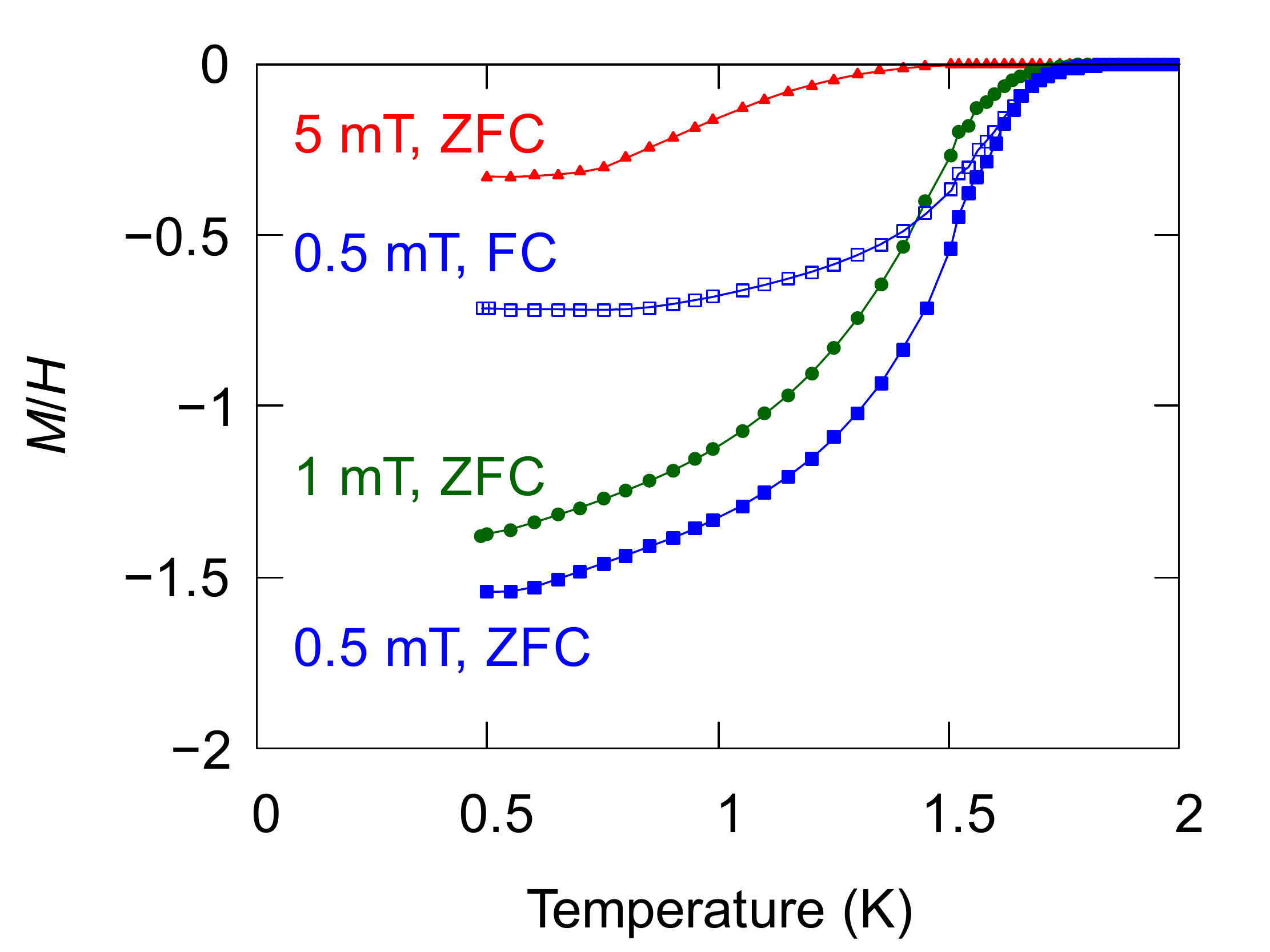}
\caption{Temperature dependence of the volumetric DC magnetic susceptibility of \ce{CaSb2}, 
measured under various fields with zero-field-cooled (ZFC) or field-cooled (FC) conditions.
The Meissner effect with the onset of 1.8~K was observed.
The shielding fraction \rev{after demagnetization correction} reaches \rev{79}\% at 0.5~K under 0.5~mT with the ZFC process.}
\label{fig: MvsT}
\end{figure}

Next, we show the DC magnetic susceptibility in Fig.~\ref{fig: MvsT}.
Strong diamagnetism indicates superconductivity below 1.8~K\@.
The \rev{susceptibility reaches $-1.54$ at 0.5~K} under 0.5~mT\@.
Assuming \rev{a rectangular} sample shape, the superconducting \rev{shielding} fraction after demagnetization correction 
is estimated to be \rev{79\%~\cite{Aharoni1998DemagFactor}}.
\rev{Later in this paper, the volume fraction is evaluated from the specific heat to be 65\%.}
The \rev{susceptibility} is suppressed to \rev{$-0.33$} under 5~mT, suggesting a lower critical field $H\sub{c1}$ smaller than 5~mT\@.

\begin{figure}
\includegraphics[width=0.85\linewidth]{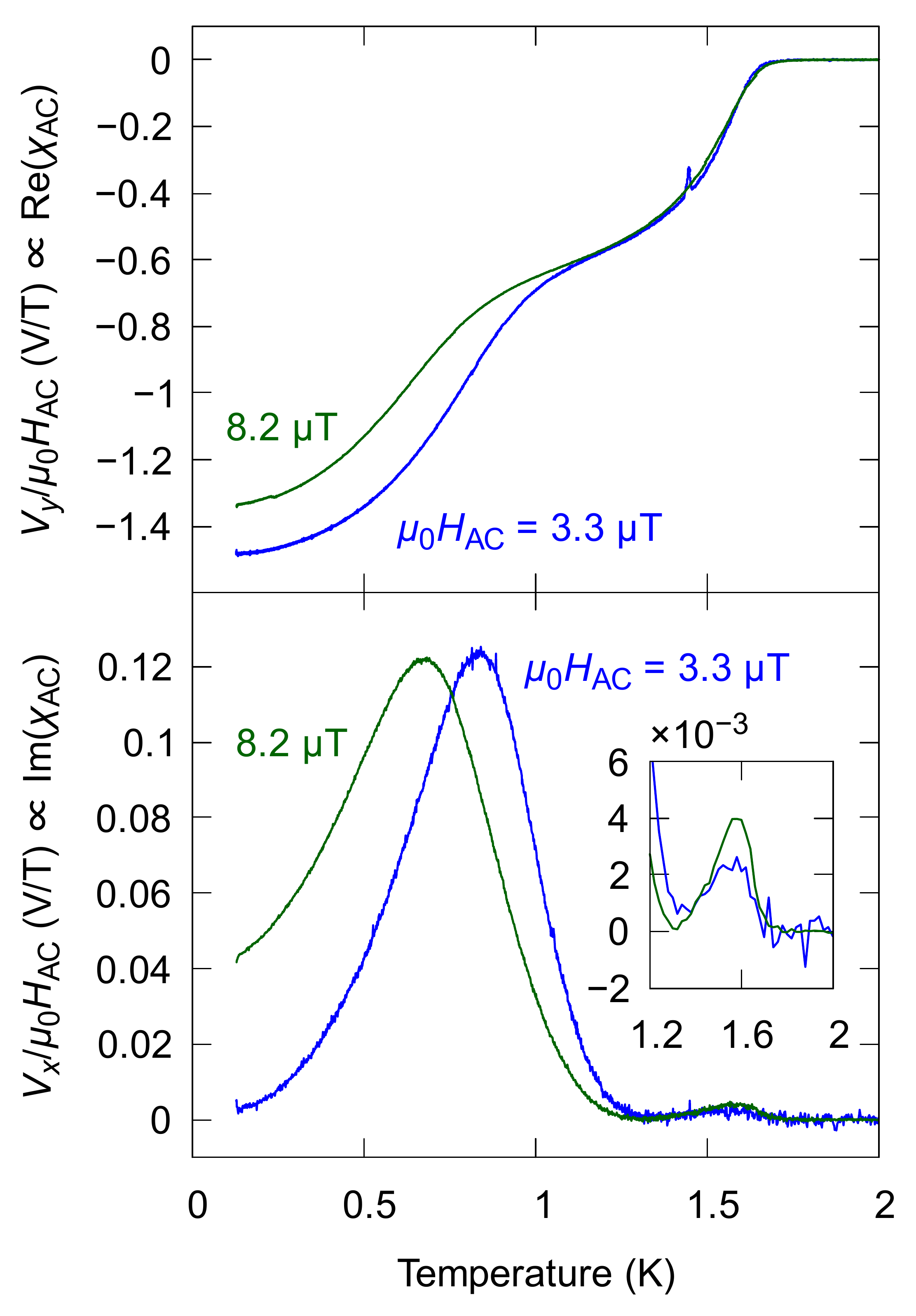}
\caption{Temperature dependence of voltage signals detected by the mutual inductance coil normalized by the AC-field amplitude.
The out-of phase signal $V_y$ and the in-phase signal $V_x$ are proportional 
to the real and imaginary parts of the AC susceptibility of \ce{CaSb2}, respectively.
These data were obtained under the AC fields with a frequency of 3011~Hz and amplitudes of 3.3 and 8.2~$\mu$T\@.
The Meissner effect was observed below 1.7~K\@.
As seen in the inset of the bottom panel, a small peak in the imaginary part was observed.
The feature at around 1~K originates from percolation of the superconducting current between the grains (see text).}
\label{fig: chiACvsT}
\end{figure}

Figure~\ref{fig: chiACvsT} shows the AC susceptibility of \ce{CaSb2}.
Diamagnetism in the real part of the susceptibility, as well as a small peak in the imaginary part, 
was observed with an onset of 1.7~K, with an AC field of 3.3 or 8.2~$\mu$T\@.
In addition to the transition at 1.7~K, we observed another broad feature at around 1~K signaled 
by an additional increase in the diamagnetic signal and by a large and broad peak in the imaginary part.
For the following three reasons, we consider that this second feature 
reflects the percolation of the superconducting current among the polycrystalline grains.
Firstly, the shielding fraction changes with the amplitude of the AC field only below the second anomaly as shown in Fig.~\ref{fig: chiACvsT}.
Secondly, the second feature is absent in the DC magnetic susceptibility as shown in Fig.~\ref{fig: MvsT}.
Thirdly, the peak of the imaginary part of the AC susceptibility is about 40 times larger at 1~K than at 1.6~K under 3.3~$\mu$T, 
even though the diamagnetism seen in the real part is comparable.
These behaviors cannot be explained by the bulk effect and are consistent with the Josephson coupling across the grain boundary.
The shielding fraction decreases with increasing AC field 
because the induced supercurrent exceeds the critical current of the grain boundary.
In addition, there is a large energy dissipation due to rapid movement of vortices along the boundary.
Similar grain-boundary effect has been reported in various polycrystalline 
or inhomogeneous materials~\cite{Yang1994GrainBoundary, Kittaka2009Sr2RuO4}.

\begin{figure}
\includegraphics[width=0.85\linewidth]{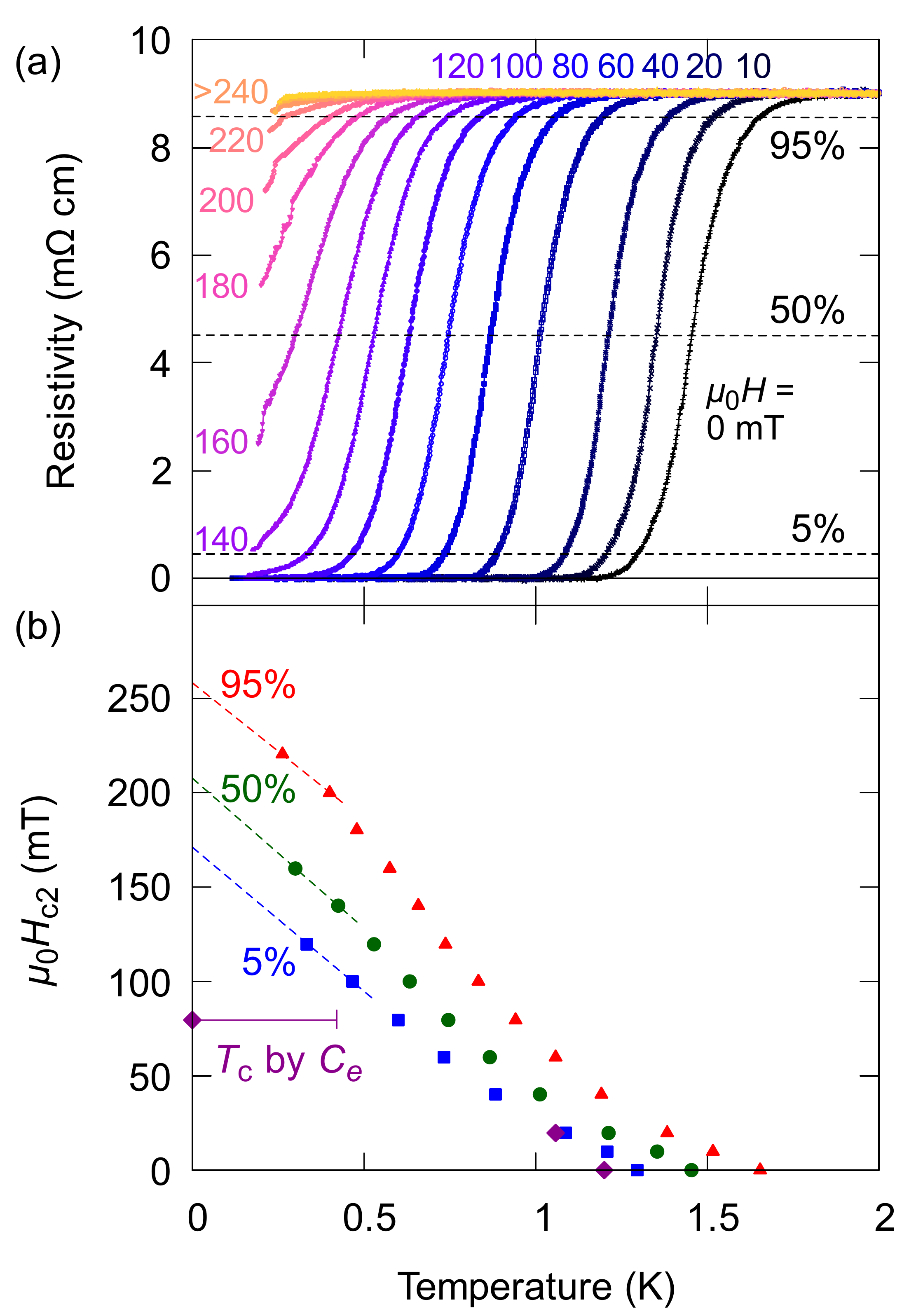}
\caption{(a) Temperature dependence of the resistivity of \ce{CaSb2} measured under 0~mT, 10~mT, and 20 to 280~mT with an interval of 20~mT\@.
Zero resistance was observed below 1.2~K under zero field.
The dashed horizontal lines present 5, 50, and 95\% of the normal-state value.
(b) Superconducting phase diagram deduced from the resistivity data.
The transition temperature under each field was defined as the temperature 
at which the resistivity becomes 5, 50, and 95\% of that in the normal state.
The dashed lines present the result of the linear fitting to the transition temperatures of each definition.
The upper critical field at 0~K was estimated to be 0.17--0.26~T\@.
\rev{Also plotted are the thermodynamic \Tc evaluated from the specific heat discussed later.}}
\label{fig: RvsT}
\end{figure}

The temperature dependence of the resistivity is presented in Fig.~\ref{fig: RvsT}(a).
Under cooling at zero field, the resistivity starts to decrease at 1.8~K and reaches zero at 1.2~K\@.
The large resistivity value of 9~m$\Omega$\,cm in the normal state is probably dominated by scatterings at the grain boundaries.
The superconductivity is suppressed as we apply the magnetic field, and above 120~mT the zero resistivity was not observed down to 0.16~K\@.
Figure~\ref{fig: RvsT}(b) shows the superconducting phase diagram deduced from the field dependence of the resistivity.
The transition temperature at each field was defined as the temperature at which the resistivity becomes 5, 50, and 95\% of the normal-state value.
As we can see especially in \Tc determined by the 95\% criteria, the slope of the phase boundary becomes steeper above 40~mT\@.
This behavior might be related to the second feature observed in the AC susceptibility (Fig.~\ref{fig: chiACvsT}).
Small applied magnetic field may break the superconductivity at the grain boundary.
As a result, the resistance increases and consequently the transition temperature rapidly decreases with small field.
At relatively low temperature, or in the high-field region in the phase diagram, 
superconductivity at the grain boundary seems less sensitive to magnetic field.
Due to this enhancement, we adopted a linear extrapolation of the upper critical field 
at the low temperatures instead of the ordinary Werthamer-Helfand-Hohenberg relation.
Then the upper critical field at 0~K is estimated to be 0.17--0.26~T depending on the definition of \Tc.
These values correspond to the Ginzburg-Landau coherence length of $\xi\sub{GL}=\sqrt{\Phi_0/(2\pi\mu_0H\sub{c2})}=36$--44~nm.
Using the information of $\mu_0H\sub{c1}<5$~mT, the penetration depth $\lambda$ and the Ginzburg-Landau parameter $\kappa=\lambda/\xi\sub{GL}$ 
are evaluated to be $\lambda>240$~nm and $\kappa>5.4$ by solving the equation 
$\mu_0H\sub{c1}=(\ln\kappa)\Phi_0/(4\pi\lambda^2)$~\cite{Tinkham2004Superconductivity}.

\begin{figure}
\includegraphics[width=0.85\linewidth]{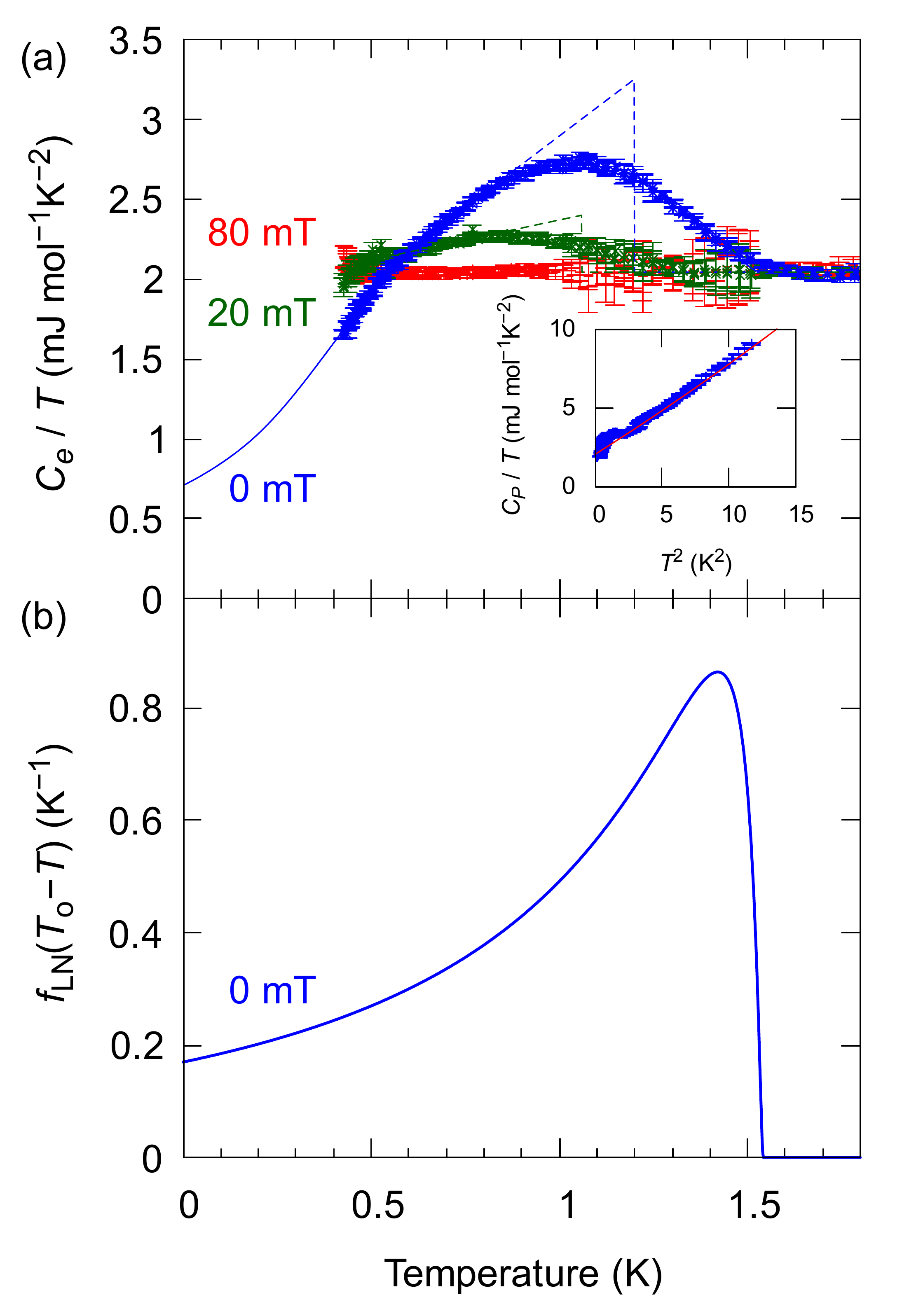}
\caption{(a) Temperature dependence of the electronic specific heat divided by temperature, evidencing the bulk superconductivity.
The solid curve presents the fitting with the BCS theory with a spatial distribution of the critical temperature.
The volume fraction is estimated to be \rev{65}\%.
\rev{The dashed lines demonstrate the estimation of the thermodynamic \Tc by entropy balance.}
The inset shows the total specific heat \rev{under 0~mT} including the phononic \rev{and Schottky} contribution\rev{s} 
divided by temperature plotted against $T^2$.
The \rev{solid} line indicates the \rev{electronic and phononic contributions} $\gamma + \beta T^2 \rev{+ \alpha T^4}$
\rev{extracted from the data under 80~mT}.
(b) The volume-fraction density $f\sub{\rev{LN}}(T\sub{o}-T)$ as a function of temperature.
Starting from $T\sub{o}=\rev{1.54}$~K, the largest population exhibits superconductivity at $T\sub{mode}=\rev{1.42}$~K 
and half of the superconductive part becomes superconducting at $T\sub{med}=\rev{1.07}$~K\@.}
\label{fig: CPvsT}
\end{figure}

Finally, we show the temperature dependence of the electronic specific heat $C_e(T)$ in Fig.~\ref{fig: CPvsT}(a).
We observed a broad but pronounced anomaly starting at 1.5~K\@.
\rev{This broadening is most likely due to inhomogeneous sample property as reflected by the AC susceptibility in Fig.~\ref{fig: chiACvsT}.}

We fitted the data \rev{under zero field} assuming 
that \Tc has a spacial distribution described by the lognormal \rev{(LN)} distribution \rev{$f\sub{LN}(x)$}:
\begin{align}
\label{eq: Cp}
C_P(T) &= C_e(T) + \beta T^3 \rev{+ \alpha T^5 + A/T^2}, \\
\frac{C_e(T)}{\gamma T} &= 1+\int_0^{T\sub{o}} \left[\frac{C\sub{BCS}(T; \Tc)}{T}-1\right]f\sub{LN}(T\sub{o}-\Tc)\mathrm{d}\Tc, \\
\label{eq: fLN}
f\sub{LN}(x) &= \frac{1}{\sqrt{2\pi}\sigma x}\exp\left(-\frac{(\ln x-\mu)^2}{2\sigma^2}\right),
\end{align}
where $C_P(T)$ is the specific heat including the phononic contribution $\beta T^3\rev{+\alpha T^5}$ \rev{and the Schottky specific heat $A/T^2$,} 
and $C\sub{BCS}(T; \Tc)$ is the theoretical specific heat \rev{of a fully gapped superconductor} derived from 
the Bardeen-Cooper-Schrieffer (BCS) theory with a critical temperature of \Tc~\cite{Tinkham2004Superconductivity}.
\rev{The Schottky specific heat likely originates from the quadrupole splitting of the Sb nucleus 
or from the Zeeman splitting of the electronic spin of possible impurities.
For $C_P$ under 20~mT, we also used $C\sub{BCS}(T;\Tc)$ valid for $H=0$.
The Sommerfeld coefficient and the phononic specific heat coefficients were determined to be
$\gamma=2.044(3)$~mJ\,mol$^{-1}$K$^{-2}$, $\beta=0.537(2)$~mJ\,mol$^{-1}$K$^{-4}$, and 
$\alpha = 0.0037(2)$~mJ\,mol$^{-1}$K$^{-6}$ from the data under $\mu_0H=80$~mT using Eq.~(\ref{eq: Cp}) with $C_e=\gamma T$.
From $\gamma$ and $\beta$, we evaluated the electronic density of states  per spin at the Fermi energy to be
$D(E\sub{F})=0.4336(7)$~eV$^{-1}$ per formula unit and the Debye temperature to be $\Theta\sub{D}=221.5(2)$~K\@.}

\rev{From the entropy balance, the thermodynamic \Tc was determined to be $\Tc=1.2$~K under 0~mT and $\Tc=1.1$~K under 20~mT\@.
First, we extrapolated the almost linear part of the experimental data to higher temperature by using a linear fitting. 
The fitting range was determined so that the root mean square of the residuals becomes the smallest. 
Then, we defined the thermodynamic \Tc as the temperature where the entropies of the extrapolation line and experimental data balance. 
As we can see in Fig.~\ref{fig: RvsT}(b), the thermodynamic \Tc is almost equal to the temperature 
at which the resistivity drops to 5\% of that in the normal state.}

Figure~\ref{fig: CPvsT}(b) presents the temperature dependence of the resultant volume-fraction density $f\sub{LN}(T\sub{o}-T)$.
\rev{From the fitting, we obtained the onset temperature $T\sub{o}=1.542(8)$~K under zero field.}
This probability distribution, restricted to $0\leq T \leq T\sub{o}$, 
has a mode of $T\sub{mode}=\rev{1.42}$~K and a median of $T\sub{med}=\rev{1.07}$~K\@.
We can see that $T\sub{med}$ \rev{coincides with} the temperature at which $C_e/T$ is maximum.
\rev{Moreover, $T\sub{o}$ and $T\sub{mode}$ roughly correspond to \Tc's determined from resistivity using the 95\% and 50\% criteria, respectively.}
\rev{The superconducting volume fraction $\int_{T}^{T\sub{o}}f\sub{\rev{LN}}(T\sub{o}-\Tc)\mathrm{d}\Tc$
is estimated to be 55\% at $T=0.5$~K and 65\% at $T=0$~K\@.}
\rev{Here, we assume that the entire part of the residual $C_e/T$ of 35\% originates 
from the normal state, as a result of the superconducting volume fraction smaller than 100\%.}
This evidences the bulk superconductivity in \ce{CaSb2}.

\section{Discussion\label{discussion}}
The first-principles calculation shows that \ce{CaSb2} hosts two \rev{sets of} Fermi surfaces 
near the zone center \rev{(attributable mainly to Sb2)} and the zone boundary \rev{(Sb1)} 
with almost the same carrier density~\cite{Funada2019CaSb2}.
Interestingly, the Fermi surface near the zone boundary \rev{consists of a pair of concentric 
cylinders}.
\rev{These cylinders along the AZ line (parallel to the $c$ axis) emerges 
from a symmetry-protected nodal line after it is slightly self-doped with electrons.
Since the nonsymmorphic space group $P2_1/m$ forces an additional orbital degeneracy in addition to the Kramers spin degeneracy,
this nodal line near the bottom of the band is four-fold degenerate~\cite{Funada2019CaSb2}.}

\rev{Due to such quartet degeneracy,} the Fermi surface \rev{consists of a pair of cylinders and} exhibits a \rev{strong mixing 
among orbitals and spins}.
\rev{Thus}, in a similar manner to doped topological insulators~\cite{Fu2010CuxBi2Se3} 
and doped Dirac semimetals~\cite{Kobayashi2015DiracSC, Kawakami2018TopoSC}, topological 
superconductivity is naturally expected if the inter-orbital pairing interaction is dominant.
In fact,the zig-zag chain\rev{s} consisting of two Sb1 sites in the unit cell 
may host such inter-orbital Cooper pairing \rev{on each cylinder}.
\rev{If the inter-orbital pairing involving inter-site Cooper pairs is realized,
an odd-parity state with nodal lines protected by the nonsymmorphic
symmetry at the Brillouin-zone boundary is expected~\cite{Kobayashi2016NonsymmorphicSC}.
Even for the intra-orbital pairing, an odd-parity state is
possible besides even-parity states, but they would be fully gapped.}

Another interesting possibility is a time-reversal-breaking chiral $p$-wave superconductivity~\cite{Shaourian2018NodalLoopSC}.
Whereas the present system has a cylindrical Fermi surface not a torus one discussed in Ref.~\cite{Shaourian2018NodalLoopSC}, 
and the relevant space group symmetry is different, it has the same feature that the Fermi surface originates from a nodal line.
Therefore, the mechanism for chiral $p$-wave superconductivity in Ref.~\cite{Shaourian2018NodalLoopSC} might work as well.

\rev{Experimental evidence for the nodal structure of the superconducting state will be given 
by the temperature-dependence of the specific heat much below \Tc.
However, the present data with a broad transition is insufficient for determination of the gap structure.
When analyzing the experimental data, we used for simplicity a theoretical curve for a fully gapped superconductivity,
which includes some of the $p$-wave superconducting states such as the Balian-Werthamer (BW) state as well as the $s$-wave states.
Determination of the superconducting gap structure using samples with a sharp transition is a challenge of future work.}

\section{Conclusion\label{conclusion}}
We report the discovery of superconductivity in \ce{CaSb2}, which has nodal lines in the bulk 
electronic band structure crossing the Fermi energy, protected by a nonsymmorphic crystalline symmetry.
Both DC magnetization and specific heat indicate the bulk superconductivity.
The transition temperature is determined to be 1.7~K from magnetic susceptibility and resistivity.
The lower critical field is smaller than 5~mT, and the upper critical field is estimated to be 0.17--0.26~T\@.
From these critical fields, the Ginzburg-Landau parameter is evaluated to be $\kappa > 5.4$.
We hope this work stimulates further investigations of the superconductivity in nonsymmorphic Dirac-line materials.


\begin{acknowledgments}
We thank the Research Center for Low Temperature and Materials Sciences in Kyoto University 
for the supply of liquid \rev{helium}.
This work was partially supported by 
by Japan Society for the Promotion of Science (JSPS) KAKENHI No.\ JP15H05851, JP15H05852, \rev{JP15H05855,} 
JP15K21717 (Topological Materials Science), 
JP17H04848, and JP17J07577, and by the JSPS Core-to-Core Program (A. Advanced Research Network), 
as well as by Izumi Science and Technology Foundation (Grant No.\ H28-J-146).
\rev{MS was supported by KAKENHI Grant No. JP17H02922 from the JSPS.}
AI is supported by the JSPS Research Fellowship.
\end{acknowledgments}


\begin{thebibliography}{40}%
\makeatletter
\providecommand \@ifxundefined [1]{%
 \@ifx{#1\undefined}
}%
\providecommand \@ifnum [1]{%
 \ifnum #1\expandafter \@firstoftwo
 \else \expandafter \@secondoftwo
 \fi
}%
\providecommand \@ifx [1]{%
 \ifx #1\expandafter \@firstoftwo
 \else \expandafter \@secondoftwo
 \fi
}%
\providecommand \natexlab [1]{#1}%
\providecommand \enquote  [1]{``#1''}%
\providecommand \bibnamefont  [1]{#1}%
\providecommand \bibfnamefont [1]{#1}%
\providecommand \citenamefont [1]{#1}%
\providecommand \href@noop [0]{\@secondoftwo}%
\providecommand \href [0]{\begingroup \@sanitize@url \@href}%
\providecommand \@href[1]{\@@startlink{#1}\@@href}%
\providecommand \@@href[1]{\endgroup#1\@@endlink}%
\providecommand \@sanitize@url [0]{\catcode `\\12\catcode `\$12\catcode
  `\&12\catcode `\#12\catcode `\^12\catcode `\_12\catcode `\%12\relax}%
\providecommand \@@startlink[1]{}%
\providecommand \@@endlink[0]{}%
\providecommand \url  [0]{\begingroup\@sanitize@url \@url }%
\providecommand \@url [1]{\endgroup\@href {#1}{\urlprefix }}%
\providecommand \urlprefix  [0]{URL }%
\providecommand \Eprint [0]{\href }%
\providecommand \doibase [0]{https://doi.org/}%
\providecommand \selectlanguage [0]{\@gobble}%
\providecommand \bibinfo  [0]{\@secondoftwo}%
\providecommand \bibfield  [0]{\@secondoftwo}%
\providecommand \translation [1]{[#1]}%
\providecommand \BibitemOpen [0]{}%
\providecommand \bibitemStop [0]{}%
\providecommand \bibitemNoStop [0]{.\EOS\space}%
\providecommand \EOS [0]{\spacefactor3000\relax}%
\providecommand \BibitemShut  [1]{\csname bibitem#1\endcsname}%
\let\auto@bib@innerbib\@empty
\bibitem [{\citenamefont {Hasan}\ and\ \citenamefont
  {Kane}(2010)}]{Hasan2010ReviewTI}%
  \BibitemOpen
  \bibfield  {author} {\bibinfo {author} {\bibfnamefont {M.~Z.}\ \bibnamefont
  {Hasan}}\ and\ \bibinfo {author} {\bibfnamefont {C.~L.}\ \bibnamefont
  {Kane}},\ }\bibfield  {title} {\bibinfo {title} {{Colloquium: Topological
  insulators}},\ }\href {https://doi.org/10.1103/RevModPhys.82.3045} {\bibfield
   {journal} {\bibinfo  {journal} {Rev. Mod. Phys.}\ }\textbf {\bibinfo
  {volume} {82}},\ \bibinfo {pages} {3045} (\bibinfo {year}
  {2010})}\BibitemShut {NoStop}%
\bibitem [{\citenamefont {Qi}\ and\ \citenamefont
  {Zhang}(2011)}]{Qi2011ReviewTIandTSC}%
  \BibitemOpen
  \bibfield  {author} {\bibinfo {author} {\bibfnamefont {X.-L.}\ \bibnamefont
  {Qi}}\ and\ \bibinfo {author} {\bibfnamefont {S.-C.}\ \bibnamefont {Zhang}},\
  }\bibfield  {title} {\bibinfo {title} {{Topological insulators and
  superconductors}},\ }\href {https://doi.org/10.1103/RevModPhys.83.1057}
  {\bibfield  {journal} {\bibinfo  {journal} {Rev. Mod. Phys.}\ }\textbf
  {\bibinfo {volume} {83}},\ \bibinfo {pages} {1057} (\bibinfo {year}
  {2011})}\BibitemShut {NoStop}%
\bibitem [{\citenamefont {Ando}(2013)}]{Ando2013ReviewTI}%
  \BibitemOpen
  \bibfield  {author} {\bibinfo {author} {\bibfnamefont {Y.}~\bibnamefont
  {Ando}},\ }\bibfield  {title} {\bibinfo {title} {{Topological Insulator
  Materials}},\ }\href {https://doi.org/10.7566/JPSJ.82.102001} {\bibfield
  {journal} {\bibinfo  {journal} {J. Phys. Soc. Jpn.}\ }\textbf {\bibinfo
  {volume} {82}},\ \bibinfo {pages} {102001} (\bibinfo {year}
  {2013})}\BibitemShut {NoStop}%
\bibitem [{\citenamefont {Tanaka}\ \emph {et~al.}(2012)\citenamefont {Tanaka},
  \citenamefont {Sato},\ and\ \citenamefont {Nagaosa}}]{Tanaka2012ReviewTSC}%
  \BibitemOpen
  \bibfield  {author} {\bibinfo {author} {\bibfnamefont {Y.}~\bibnamefont
  {Tanaka}}, \bibinfo {author} {\bibfnamefont {M.}~\bibnamefont {Sato}},\ and\
  \bibinfo {author} {\bibfnamefont {N.}~\bibnamefont {Nagaosa}},\ }\bibfield
  {title} {\bibinfo {title} {{Symmetry and Topology in Superconductors
  --Odd-Frequency Pairing and Edge States--}},\ }\href
  {https://doi.org/10.1143/JPSJ.81.011013} {\bibfield  {journal} {\bibinfo
  {journal} {J. Phys. Soc. Jpn.}\ }\textbf {\bibinfo {volume} {81}},\ \bibinfo
  {pages} {011013} (\bibinfo {year} {2012})}\BibitemShut {NoStop}%
\bibitem [{\citenamefont {Sato}\ and\ \citenamefont
  {Fujimoto}(2016)}]{Sato2016ReviewTSC}%
  \BibitemOpen
  \bibfield  {author} {\bibinfo {author} {\bibfnamefont {M.}~\bibnamefont
  {Sato}}\ and\ \bibinfo {author} {\bibfnamefont {S.}~\bibnamefont
  {Fujimoto}},\ }\bibfield  {title} {\bibinfo {title} {{Majorana Fermions and
  Topology in Superconductors}},\ }\href
  {https://doi.org/10.7566/JPSJ.85.072001} {\bibfield  {journal} {\bibinfo
  {journal} {J. Phys. Soc. Jpn.}\ }\textbf {\bibinfo {volume} {85}},\ \bibinfo
  {pages} {072001} (\bibinfo {year} {2016})}\BibitemShut {NoStop}%
\bibitem [{\citenamefont {Sato}\ and\ \citenamefont
  {Ando}(2017)}]{Sato2017ReviewTSC}%
  \BibitemOpen
  \bibfield  {author} {\bibinfo {author} {\bibfnamefont {M.}~\bibnamefont
  {Sato}}\ and\ \bibinfo {author} {\bibfnamefont {Y.}~\bibnamefont {Ando}},\
  }\bibfield  {title} {\bibinfo {title} {{Topological superconductors: a
  review}},\ }\href {https://doi.org/10.1088/1361-6633/aa6ac7} {\bibfield
  {journal} {\bibinfo  {journal} {Reports on Progress in Physics}\ }\textbf
  {\bibinfo {volume} {80}},\ \bibinfo {pages} {076501} (\bibinfo {year}
  {2017})}\BibitemShut {NoStop}%
\bibitem [{\citenamefont {Fu}\ and\ \citenamefont
  {Berg}(2010)}]{Fu2010CuxBi2Se3}%
  \BibitemOpen
  \bibfield  {author} {\bibinfo {author} {\bibfnamefont {L.}~\bibnamefont
  {Fu}}\ and\ \bibinfo {author} {\bibfnamefont {E.}~\bibnamefont {Berg}},\
  }\bibfield  {title} {\bibinfo {title} {{Odd-Parity Topological
  Superconductors: Theory and Application to \ce{Cu_xBi2Se3}}},\ }\href
  {https://doi.org/10.1103/PhysRevLett.105.097001} {\bibfield  {journal}
  {\bibinfo  {journal} {Phys. Rev. Lett.}\ }\textbf {\bibinfo {volume} {105}},\
  \bibinfo {pages} {097001} (\bibinfo {year} {2010})}\BibitemShut {NoStop}%
\bibitem [{\citenamefont {Sasaki}\ \emph {et~al.}(2011)\citenamefont {Sasaki},
  \citenamefont {Kriener}, \citenamefont {Segawa}, \citenamefont {Yada},
  \citenamefont {Tanaka}, \citenamefont {Sato},\ and\ \citenamefont
  {Ando}}]{Sasaki2011CuxBi2Se3}%
  \BibitemOpen
  \bibfield  {author} {\bibinfo {author} {\bibfnamefont {S.}~\bibnamefont
  {Sasaki}}, \bibinfo {author} {\bibfnamefont {M.}~\bibnamefont {Kriener}},
  \bibinfo {author} {\bibfnamefont {K.}~\bibnamefont {Segawa}}, \bibinfo
  {author} {\bibfnamefont {K.}~\bibnamefont {Yada}}, \bibinfo {author}
  {\bibfnamefont {Y.}~\bibnamefont {Tanaka}}, \bibinfo {author} {\bibfnamefont
  {M.}~\bibnamefont {Sato}},\ and\ \bibinfo {author} {\bibfnamefont
  {Y.}~\bibnamefont {Ando}},\ }\bibfield  {title} {\bibinfo {title}
  {{Topological Superconductivity in \ce{Cu_xBi2Se3}}},\ }\href
  {https://doi.org/10.1103/PhysRevLett.107.217001} {\bibfield  {journal}
  {\bibinfo  {journal} {Phys. Rev. Lett.}\ }\textbf {\bibinfo {volume} {107}},\
  \bibinfo {pages} {217001} (\bibinfo {year} {2011})}\BibitemShut {NoStop}%
\bibitem [{\citenamefont {Matano}\ \emph {et~al.}(2016)\citenamefont {Matano},
  \citenamefont {Kriener}, \citenamefont {Segawa}, \citenamefont {Ando},\ and\
  \citenamefont {Zheng}}]{Matano2016CuxBi2Se3}%
  \BibitemOpen
  \bibfield  {author} {\bibinfo {author} {\bibfnamefont {K.}~\bibnamefont
  {Matano}}, \bibinfo {author} {\bibfnamefont {M.}~\bibnamefont {Kriener}},
  \bibinfo {author} {\bibfnamefont {K.}~\bibnamefont {Segawa}}, \bibinfo
  {author} {\bibfnamefont {Y.}~\bibnamefont {Ando}},\ and\ \bibinfo {author}
  {\bibfnamefont {G.-q.}\ \bibnamefont {Zheng}},\ }\bibfield  {title} {\bibinfo
  {title} {{Spin-rotation symmetry breaking in the superconducting state of
  \ce{Cu_xBi2Se3}}},\ }\href {https://doi.org/10.1038/nphys3781} {\bibfield
  {journal} {\bibinfo  {journal} {Nat. Phys.}\ }\textbf {\bibinfo {volume}
  {12}},\ \bibinfo {pages} {852} (\bibinfo {year} {2016})}\BibitemShut
  {NoStop}%
\bibitem [{\citenamefont {Yonezawa}\ \emph {et~al.}(2017)\citenamefont
  {Yonezawa}, \citenamefont {Tajiri}, \citenamefont {Nakata}, \citenamefont
  {Nagai}, \citenamefont {Wang}, \citenamefont {Segawa}, \citenamefont {Ando},\
  and\ \citenamefont {Maeno}}]{Tajiri2017CuxBi2Se3}%
  \BibitemOpen
  \bibfield  {author} {\bibinfo {author} {\bibfnamefont {S.}~\bibnamefont
  {Yonezawa}}, \bibinfo {author} {\bibfnamefont {K.}~\bibnamefont {Tajiri}},
  \bibinfo {author} {\bibfnamefont {S.}~\bibnamefont {Nakata}}, \bibinfo
  {author} {\bibfnamefont {Y.}~\bibnamefont {Nagai}}, \bibinfo {author}
  {\bibfnamefont {Z.}~\bibnamefont {Wang}}, \bibinfo {author} {\bibfnamefont
  {K.}~\bibnamefont {Segawa}}, \bibinfo {author} {\bibfnamefont
  {Y.}~\bibnamefont {Ando}},\ and\ \bibinfo {author} {\bibfnamefont
  {Y.}~\bibnamefont {Maeno}},\ }\bibfield  {title} {\bibinfo {title}
  {{Thermodynamic evidence for nematic superconductivity in \ce{Cu_xBi2Se3}}},\
  }\href {https://doi.org/10.1038/nphys3907} {\bibfield  {journal} {\bibinfo
  {journal} {Nat. Phys.}\ }\textbf {\bibinfo {volume} {13}},\ \bibinfo {pages}
  {123} (\bibinfo {year} {2017})}\BibitemShut {NoStop}%
\bibitem [{\citenamefont {Yan}\ and\ \citenamefont
  {Felser}(2017)}]{Yan2017ReviewWeyl}%
  \BibitemOpen
  \bibfield  {author} {\bibinfo {author} {\bibfnamefont {B.}~\bibnamefont
  {Yan}}\ and\ \bibinfo {author} {\bibfnamefont {C.}~\bibnamefont {Felser}},\
  }\bibfield  {title} {\bibinfo {title} {{Topological Materials: Weyl
  Semimetals}},\ }\href
  {https://doi.org/10.1146/annurev-conmatphys-031016-025458} {\bibfield
  {journal} {\bibinfo  {journal} {Annu. Rev. Condens. Matter. Phys.}\ }\textbf
  {\bibinfo {volume} {8}},\ \bibinfo {pages} {337} (\bibinfo {year}
  {2017})}\BibitemShut {NoStop}%
\bibitem [{\citenamefont {Armitage}\ \emph {et~al.}(2018)\citenamefont
  {Armitage}, \citenamefont {Mele},\ and\ \citenamefont
  {Vishwanath}}]{Armitage2018ReviewDiracWeyl}%
  \BibitemOpen
  \bibfield  {author} {\bibinfo {author} {\bibfnamefont {N.~P.}\ \bibnamefont
  {Armitage}}, \bibinfo {author} {\bibfnamefont {E.~J.}\ \bibnamefont {Mele}},\
  and\ \bibinfo {author} {\bibfnamefont {A.}~\bibnamefont {Vishwanath}},\
  }\bibfield  {title} {\bibinfo {title} {{Weyl and Dirac semimetals in
  three-dimensional solids}},\ }\href
  {https://doi.org/10.1103/RevModPhys.90.015001} {\bibfield  {journal}
  {\bibinfo  {journal} {Rev. Mod. Phys.}\ }\textbf {\bibinfo {volume} {90}},\
  \bibinfo {pages} {015001} (\bibinfo {year} {2018})}\BibitemShut {NoStop}%
\bibitem [{\citenamefont {Kobayashi}\ and\ \citenamefont
  {Sato}(2015)}]{Kobayashi2015DiracSC}%
  \BibitemOpen
  \bibfield  {author} {\bibinfo {author} {\bibfnamefont {S.}~\bibnamefont
  {Kobayashi}}\ and\ \bibinfo {author} {\bibfnamefont {M.}~\bibnamefont
  {Sato}},\ }\bibfield  {title} {\bibinfo {title} {{Topological
  Superconductivity in Dirac Semimetals}},\ }\href
  {https://doi.org/10.1103/PhysRevLett.115.187001} {\bibfield  {journal}
  {\bibinfo  {journal} {Phys. Rev. Lett.}\ }\textbf {\bibinfo {volume} {115}},\
  \bibinfo {pages} {187001} (\bibinfo {year} {2015})}\BibitemShut {NoStop}%
\bibitem [{\citenamefont {Li}\ and\ \citenamefont
  {Haldane}(2018)}]{Li2018DiracSC}%
  \BibitemOpen
  \bibfield  {author} {\bibinfo {author} {\bibfnamefont {Y.}~\bibnamefont
  {Li}}\ and\ \bibinfo {author} {\bibfnamefont {F.~D.~M.}\ \bibnamefont
  {Haldane}},\ }\bibfield  {title} {\bibinfo {title} {{Topological Nodal Cooper
  Pairing in Doped Weyl Metals}},\ }\href
  {https://doi.org/10.1103/PhysRevLett.120.067003} {\bibfield  {journal}
  {\bibinfo  {journal} {Phys. Rev. Lett.}\ }\textbf {\bibinfo {volume} {120}},\
  \bibinfo {pages} {067003} (\bibinfo {year} {2018})}\BibitemShut {NoStop}%
\bibitem [{\citenamefont {Kawakami}\ \emph {et~al.}(2018)\citenamefont
  {Kawakami}, \citenamefont {Okamura}, \citenamefont {Kobayashi},\ and\
  \citenamefont {Sato}}]{Kawakami2018TopoSC}%
  \BibitemOpen
  \bibfield  {author} {\bibinfo {author} {\bibfnamefont {T.}~\bibnamefont
  {Kawakami}}, \bibinfo {author} {\bibfnamefont {T.}~\bibnamefont {Okamura}},
  \bibinfo {author} {\bibfnamefont {S.}~\bibnamefont {Kobayashi}},\ and\
  \bibinfo {author} {\bibfnamefont {M.}~\bibnamefont {Sato}},\ }\bibfield
  {title} {\bibinfo {title} {{Topological Crystalline Materials of $J=3/2$
  Electrons: Antiperovskites, Dirac Points, and High Winding Topological
  Superconductivity}},\ }\href {https://doi.org/10.1103/PhysRevX.8.041026}
  {\bibfield  {journal} {\bibinfo  {journal} {Phys. Rev. X}\ }\textbf {\bibinfo
  {volume} {8}},\ \bibinfo {pages} {041026} (\bibinfo {year}
  {2018})}\BibitemShut {NoStop}%
\bibitem [{\citenamefont {He}\ \emph {et~al.}(2016)\citenamefont {He},
  \citenamefont {Jia}, \citenamefont {Zhang}, \citenamefont {Hong},
  \citenamefont {Jin},\ and\ \citenamefont {Li}}]{He2016Cd3As2}%
  \BibitemOpen
  \bibfield  {author} {\bibinfo {author} {\bibfnamefont {L.}~\bibnamefont
  {He}}, \bibinfo {author} {\bibfnamefont {Y.}~\bibnamefont {Jia}}, \bibinfo
  {author} {\bibfnamefont {S.}~\bibnamefont {Zhang}}, \bibinfo {author}
  {\bibfnamefont {X.}~\bibnamefont {Hong}}, \bibinfo {author} {\bibfnamefont
  {C.}~\bibnamefont {Jin}},\ and\ \bibinfo {author} {\bibfnamefont
  {S.}~\bibnamefont {Li}},\ }\bibfield  {title} {\bibinfo {title}
  {{Pressure-induced superconductivity in the three-dimensional topological
  Dirac semimetal \ce{Cd3As2}}},\ }\href
  {https://doi.org/10.1038/npjquantmats.2016.14} {\bibfield  {journal}
  {\bibinfo  {journal} {npj Quantum Mater.}\ }\textbf {\bibinfo {volume} {1}},\
  \bibinfo {pages} {16014} (\bibinfo {year} {2016})}\BibitemShut {NoStop}%
\bibitem [{\citenamefont {Oudah}\ \emph {et~al.}(2016)\citenamefont {Oudah},
  \citenamefont {Ikeda}, \citenamefont {Hausmann}, \citenamefont {Yonezawa},
  \citenamefont {Fukumoto}, \citenamefont {Kobayashi}, \citenamefont {Sato},\
  and\ \citenamefont {Maeno}}]{Oudah2016superconductivity}%
  \BibitemOpen
  \bibfield  {author} {\bibinfo {author} {\bibfnamefont {M.}~\bibnamefont
  {Oudah}}, \bibinfo {author} {\bibfnamefont {A.}~\bibnamefont {Ikeda}},
  \bibinfo {author} {\bibfnamefont {J.~N.}\ \bibnamefont {Hausmann}}, \bibinfo
  {author} {\bibfnamefont {S.}~\bibnamefont {Yonezawa}}, \bibinfo {author}
  {\bibfnamefont {T.}~\bibnamefont {Fukumoto}}, \bibinfo {author}
  {\bibfnamefont {S.}~\bibnamefont {Kobayashi}}, \bibinfo {author}
  {\bibfnamefont {M.}~\bibnamefont {Sato}},\ and\ \bibinfo {author}
  {\bibfnamefont {Y.}~\bibnamefont {Maeno}},\ }\bibfield  {title} {\bibinfo
  {title} {{Superconductivity in the antiperovskite Dirac-metal oxide
  \ce{Sr_{3-x}SnO}}},\ }\href {https://doi.org/10.1038/ncomms13617} {\bibfield
  {journal} {\bibinfo  {journal} {Nat. Commun.}\ }\textbf {\bibinfo {volume}
  {7}},\ \bibinfo {pages} {13617} (\bibinfo {year} {2016})}\BibitemShut
  {NoStop}%
\bibitem [{\citenamefont {Fang}\ \emph {et~al.}(2015)\citenamefont {Fang},
  \citenamefont {Chen}, \citenamefont {Kee},\ and\ \citenamefont
  {Fu}}]{Fang2015NodalLine}%
  \BibitemOpen
  \bibfield  {author} {\bibinfo {author} {\bibfnamefont {C.}~\bibnamefont
  {Fang}}, \bibinfo {author} {\bibfnamefont {Y.}~\bibnamefont {Chen}}, \bibinfo
  {author} {\bibfnamefont {H.-Y.}\ \bibnamefont {Kee}},\ and\ \bibinfo {author}
  {\bibfnamefont {L.}~\bibnamefont {Fu}},\ }\bibfield  {title} {\bibinfo
  {title} {{Topological nodal line semimetals with and without spin-orbital
  coupling}},\ }\href {https://doi.org/10.1103/PhysRevB.92.081201} {\bibfield
  {journal} {\bibinfo  {journal} {Phys. Rev. B}\ }\textbf {\bibinfo {volume}
  {92}},\ \bibinfo {pages} {081201(R)} (\bibinfo {year} {2015})}\BibitemShut
  {NoStop}%
\bibitem [{\citenamefont {Yamakage}\ \emph {et~al.}(2016)\citenamefont
  {Yamakage}, \citenamefont {Yamakawa}, \citenamefont {Tanaka},\ and\
  \citenamefont {Okamoto}}]{Yamakage2016CaAgX}%
  \BibitemOpen
  \bibfield  {author} {\bibinfo {author} {\bibfnamefont {A.}~\bibnamefont
  {Yamakage}}, \bibinfo {author} {\bibfnamefont {Y.}~\bibnamefont {Yamakawa}},
  \bibinfo {author} {\bibfnamefont {Y.}~\bibnamefont {Tanaka}},\ and\ \bibinfo
  {author} {\bibfnamefont {Y.}~\bibnamefont {Okamoto}},\ }\bibfield  {title}
  {\bibinfo {title} {{Line-Node Dirac Semimetal and Topological Insulating
  Phase in Noncentrosymmetric Pnictides CaAg$X$ ($X$ = P, As)}},\ }\href
  {https://doi.org/10.7566/JPSJ.85.013708} {\bibfield  {journal} {\bibinfo
  {journal} {J. Phys. Soc. Jpn.}\ }\textbf {\bibinfo {volume} {85}},\ \bibinfo
  {pages} {013708} (\bibinfo {year} {2016})}\BibitemShut {NoStop}%
\bibitem [{\citenamefont {Kobayashi}\ \emph {et~al.}(2017)\citenamefont
  {Kobayashi}, \citenamefont {Yamakawa}, \citenamefont {Yamakage},
  \citenamefont {Inohara}, \citenamefont {Okamoto},\ and\ \citenamefont
  {Tanaka}}]{Kobayashi2017NodalLine}%
  \BibitemOpen
  \bibfield  {author} {\bibinfo {author} {\bibfnamefont {S.}~\bibnamefont
  {Kobayashi}}, \bibinfo {author} {\bibfnamefont {Y.}~\bibnamefont {Yamakawa}},
  \bibinfo {author} {\bibfnamefont {A.}~\bibnamefont {Yamakage}}, \bibinfo
  {author} {\bibfnamefont {T.}~\bibnamefont {Inohara}}, \bibinfo {author}
  {\bibfnamefont {Y.}~\bibnamefont {Okamoto}},\ and\ \bibinfo {author}
  {\bibfnamefont {Y.}~\bibnamefont {Tanaka}},\ }\bibfield  {title} {\bibinfo
  {title} {{Crossing-line-node semimetals: General theory and application to
  rare-earth trihydrides}},\ }\href
  {https://doi.org/10.1103/PhysRevB.95.245208} {\bibfield  {journal} {\bibinfo
  {journal} {Phys. Rev. B}\ }\textbf {\bibinfo {volume} {95}},\ \bibinfo
  {pages} {245208} (\bibinfo {year} {2017})}\BibitemShut {NoStop}%
\bibitem [{\citenamefont {Huh}\ \emph {et~al.}(2016)\citenamefont {Huh},
  \citenamefont {Moon},\ and\ \citenamefont {Kim}}]{Huh2016NodalRing}%
  \BibitemOpen
  \bibfield  {author} {\bibinfo {author} {\bibfnamefont {Y.}~\bibnamefont
  {Huh}}, \bibinfo {author} {\bibfnamefont {E.-G.}\ \bibnamefont {Moon}},\ and\
  \bibinfo {author} {\bibfnamefont {Y.~B.}\ \bibnamefont {Kim}},\ }\bibfield
  {title} {\bibinfo {title} {{Long-range Coulomb interaction in nodal-ring
  semimetals}},\ }\href {https://doi.org/10.1103/PhysRevB.93.035138} {\bibfield
   {journal} {\bibinfo  {journal} {Phys. Rev. B}\ }\textbf {\bibinfo {volume}
  {93}},\ \bibinfo {pages} {035138} (\bibinfo {year} {2016})}\BibitemShut
  {NoStop}%
\bibitem [{\citenamefont {Hirayama}\ \emph {et~al.}(2017)\citenamefont
  {Hirayama}, \citenamefont {Okugawa}, \citenamefont {Miyake},\ and\
  \citenamefont {Murakami}}]{Hirayama2017NodalLine}%
  \BibitemOpen
  \bibfield  {author} {\bibinfo {author} {\bibfnamefont {M.}~\bibnamefont
  {Hirayama}}, \bibinfo {author} {\bibfnamefont {R.}~\bibnamefont {Okugawa}},
  \bibinfo {author} {\bibfnamefont {T.}~\bibnamefont {Miyake}},\ and\ \bibinfo
  {author} {\bibfnamefont {S.}~\bibnamefont {Murakami}},\ }\bibfield  {title}
  {\bibinfo {title} {{Topological Dirac nodal lines and surface charges in fcc
  alkaline earth metals}},\ }\href {https://doi.org/10.1038/ncomms14022}
  {\bibfield  {journal} {\bibinfo  {journal} {Nat. Commun.}\ }\textbf {\bibinfo
  {volume} {8}},\ \bibinfo {pages} {14022} (\bibinfo {year}
  {2017})}\BibitemShut {NoStop}%
\bibitem [{\citenamefont {Ramamurthy}\ and\ \citenamefont
  {Hughes}(2017)}]{Ramamurthy2017NodalLine}%
  \BibitemOpen
  \bibfield  {author} {\bibinfo {author} {\bibfnamefont {S.~T.}\ \bibnamefont
  {Ramamurthy}}\ and\ \bibinfo {author} {\bibfnamefont {T.~L.}\ \bibnamefont
  {Hughes}},\ }\bibfield  {title} {\bibinfo {title} {{Quasitopological
  electromagnetic response of line-node semimetals}},\ }\href
  {https://doi.org/10.1103/PhysRevB.95.075138} {\bibfield  {journal} {\bibinfo
  {journal} {Phys. Rev. B}\ }\textbf {\bibinfo {volume} {95}},\ \bibinfo
  {pages} {075138} (\bibinfo {year} {2017})}\BibitemShut {NoStop}%
\bibitem [{\citenamefont {Shapourian}\ \emph {et~al.}(2018)\citenamefont
  {Shapourian}, \citenamefont {Wang},\ and\ \citenamefont
  {Ryu}}]{Shaourian2018NodalLoopSC}%
  \BibitemOpen
  \bibfield  {author} {\bibinfo {author} {\bibfnamefont {H.}~\bibnamefont
  {Shapourian}}, \bibinfo {author} {\bibfnamefont {Y.}~\bibnamefont {Wang}},\
  and\ \bibinfo {author} {\bibfnamefont {S.}~\bibnamefont {Ryu}},\ }\bibfield
  {title} {\bibinfo {title} {{Topological crystalline superconductivity and
  second-order topological superconductivity in nodal-loop materials}},\ }\href
  {https://doi.org/10.1103/PhysRevB.97.094508} {\bibfield  {journal} {\bibinfo
  {journal} {Phys. Rev. B}\ }\textbf {\bibinfo {volume} {97}},\ \bibinfo
  {pages} {094508} (\bibinfo {year} {2018})}\BibitemShut {NoStop}%
\bibitem [{\citenamefont {Ali}\ \emph {et~al.}(2014)\citenamefont {Ali},
  \citenamefont {Gibson}, \citenamefont {Klimczuk},\ and\ \citenamefont
  {Cava}}]{Ali2014PbTaSe2}%
  \BibitemOpen
  \bibfield  {author} {\bibinfo {author} {\bibfnamefont {M.~N.}\ \bibnamefont
  {Ali}}, \bibinfo {author} {\bibfnamefont {Q.~D.}\ \bibnamefont {Gibson}},
  \bibinfo {author} {\bibfnamefont {T.}~\bibnamefont {Klimczuk}},\ and\
  \bibinfo {author} {\bibfnamefont {R.~J.}\ \bibnamefont {Cava}},\ }\bibfield
  {title} {\bibinfo {title} {{Noncentrosymmetric superconductor with a bulk
  three-dimensional Dirac cone gapped by strong spin-orbit coupling}},\ }\href
  {https://doi.org/10.1103/PhysRevB.89.020505} {\bibfield  {journal} {\bibinfo
  {journal} {Phys. Rev. B}\ }\textbf {\bibinfo {volume} {89}},\ \bibinfo
  {pages} {020505(R)} (\bibinfo {year} {2014})}\BibitemShut {NoStop}%
\bibitem [{\citenamefont {Carter}\ \emph {et~al.}(2012)\citenamefont {Carter},
  \citenamefont {Shankar}, \citenamefont {Zeb},\ and\ \citenamefont
  {Kee}}]{Carter2012NodalLine}%
  \BibitemOpen
  \bibfield  {author} {\bibinfo {author} {\bibfnamefont {J.-M.}\ \bibnamefont
  {Carter}}, \bibinfo {author} {\bibfnamefont {V.~V.}\ \bibnamefont {Shankar}},
  \bibinfo {author} {\bibfnamefont {M.~A.}\ \bibnamefont {Zeb}},\ and\ \bibinfo
  {author} {\bibfnamefont {H.-Y.}\ \bibnamefont {Kee}},\ }\bibfield  {title}
  {\bibinfo {title} {Semimetal and topological insulator in perovskite
  iridates},\ }\href {https://doi.org/10.1103/PhysRevB.85.115105} {\bibfield
  {journal} {\bibinfo  {journal} {Phys. Rev. B}\ }\textbf {\bibinfo {volume}
  {85}},\ \bibinfo {pages} {115105} (\bibinfo {year} {2012})}\BibitemShut
  {NoStop}%
\bibitem [{\citenamefont {Chen}\ \emph {et~al.}(2015)\citenamefont {Chen},
  \citenamefont {Lu},\ and\ \citenamefont {Kee}}]{Chen2015NodalLine}%
  \BibitemOpen
  \bibfield  {author} {\bibinfo {author} {\bibfnamefont {Y.}~\bibnamefont
  {Chen}}, \bibinfo {author} {\bibfnamefont {Y.-M.}\ \bibnamefont {Lu}},\ and\
  \bibinfo {author} {\bibfnamefont {H.-Y.}\ \bibnamefont {Kee}},\ }\bibfield
  {title} {\bibinfo {title} {{Topological crystalline metal in orthorhombic
  perovskite iridates}},\ }\href {https://doi.org/10.1038/ncomms7593}
  {\bibfield  {journal} {\bibinfo  {journal} {Nat. Commun.}\ }\textbf {\bibinfo
  {volume} {6}},\ \bibinfo {pages} {6593} (\bibinfo {year} {2015})}\BibitemShut
  {NoStop}%
\bibitem [{\citenamefont {Liang}\ \emph {et~al.}(2016)\citenamefont {Liang},
  \citenamefont {Zhou}, \citenamefont {Yu}, \citenamefont {Wang},\ and\
  \citenamefont {Weng}}]{Liang2016NodalLine}%
  \BibitemOpen
  \bibfield  {author} {\bibinfo {author} {\bibfnamefont {Q.-F.}\ \bibnamefont
  {Liang}}, \bibinfo {author} {\bibfnamefont {J.}~\bibnamefont {Zhou}},
  \bibinfo {author} {\bibfnamefont {R.}~\bibnamefont {Yu}}, \bibinfo {author}
  {\bibfnamefont {Z.}~\bibnamefont {Wang}},\ and\ \bibinfo {author}
  {\bibfnamefont {H.}~\bibnamefont {Weng}},\ }\bibfield  {title} {\bibinfo
  {title} {{Node-surface and node-line fermions from nonsymmorphic lattice
  symmetries}},\ }\href {https://doi.org/10.1103/PhysRevB.93.085427} {\bibfield
   {journal} {\bibinfo  {journal} {Phys. Rev. B}\ }\textbf {\bibinfo {volume}
  {93}},\ \bibinfo {pages} {085427} (\bibinfo {year} {2016})}\BibitemShut
  {NoStop}%
\bibitem [{\citenamefont {Sun}\ \emph {et~al.}(2017)\citenamefont {Sun},
  \citenamefont {Zhang}, \citenamefont {Liu}, \citenamefont {Felser},\ and\
  \citenamefont {Yan}}]{Sun2017NodalLine}%
  \BibitemOpen
  \bibfield  {author} {\bibinfo {author} {\bibfnamefont {Y.}~\bibnamefont
  {Sun}}, \bibinfo {author} {\bibfnamefont {Y.}~\bibnamefont {Zhang}}, \bibinfo
  {author} {\bibfnamefont {C.-X.}\ \bibnamefont {Liu}}, \bibinfo {author}
  {\bibfnamefont {C.}~\bibnamefont {Felser}},\ and\ \bibinfo {author}
  {\bibfnamefont {B.}~\bibnamefont {Yan}},\ }\bibfield  {title} {\bibinfo
  {title} {{Dirac nodal lines and induced spin Hall effect in metallic rutile
  oxides}},\ }\href {https://doi.org/10.1103/PhysRevB.95.235104} {\bibfield
  {journal} {\bibinfo  {journal} {Phys. Rev. B}\ }\textbf {\bibinfo {volume}
  {95}},\ \bibinfo {pages} {235104} (\bibinfo {year} {2017})}\BibitemShut
  {NoStop}%
\bibitem [{\citenamefont {Neumeier}\ and\ \citenamefont
  {Smith}(2017)}]{Neumeier2017BaSbS3}%
  \BibitemOpen
  \bibfield  {author} {\bibinfo {author} {\bibfnamefont {J.~J.}\ \bibnamefont
  {Neumeier}}\ and\ \bibinfo {author} {\bibfnamefont {M.~G.}\ \bibnamefont
  {Smith}},\ }\bibfield  {title} {\bibinfo {title} {{Superconductivity in
  quasi-one-dimensional \ce{BaNbS3}}},\ }\href
  {https://doi.org/https://doi.org/10.1016/j.physc.2017.08.006} {\bibfield
  {journal} {\bibinfo  {journal} {Physica C}\ }\textbf {\bibinfo {volume}
  {542}},\ \bibinfo {pages} {1} (\bibinfo {year} {2017})}\BibitemShut {NoStop}%
\bibitem [{\citenamefont {Funada}\ \emph {et~al.}(2019)\citenamefont {Funada},
  \citenamefont {Yamakage}, \citenamefont {Yamashina},\ and\ \citenamefont
  {Kageyama}}]{Funada2019CaSb2}%
  \BibitemOpen
  \bibfield  {author} {\bibinfo {author} {\bibfnamefont {K.}~\bibnamefont
  {Funada}}, \bibinfo {author} {\bibfnamefont {A.}~\bibnamefont {Yamakage}},
  \bibinfo {author} {\bibfnamefont {N.}~\bibnamefont {Yamashina}},\ and\
  \bibinfo {author} {\bibfnamefont {H.}~\bibnamefont {Kageyama}},\ }\bibfield
  {title} {\bibinfo {title} {{Spin-Orbit-Coupling-Induced Type-I/type-II Dirac
  Nodal-Line Metal in Nonsymmorphic \ce{CaSb2}}},\ }\href
  {https://doi.org/10.7566/JPSJ.88.044711} {\bibfield  {journal} {\bibinfo
  {journal} {J. Phys. Soc. Jpn.}\ }\textbf {\bibinfo {volume} {88}},\ \bibinfo
  {pages} {044711} (\bibinfo {year} {2019})}\BibitemShut {NoStop}%
\bibitem [{\citenamefont {Deller}\ and\ \citenamefont
  {Eisenmann}(1976)}]{Deller1976CaSb2}%
  \BibitemOpen
  \bibfield  {author} {\bibinfo {author} {\bibfnamefont {K.}~\bibnamefont
  {Deller}}\ and\ \bibinfo {author} {\bibfnamefont {B.}~\bibnamefont
  {Eisenmann}},\ }\bibfield  {title} {\bibinfo {title} {{Darstellung und
  Kristallstruktur von \ce{CaSb2}}},\ }\href
  {https://doi.org/10.1002/zaac.19764250203} {\bibfield  {journal} {\bibinfo
  {journal} {Z. Anorg. Allg. Chem.}\ }\textbf {\bibinfo {volume} {425}},\
  \bibinfo {pages} {104} (\bibinfo {year} {1976})}\BibitemShut {NoStop}%
\bibitem [{\citenamefont {Momma}\ and\ \citenamefont
  {Izumi}(2011)}]{Momma2011VESTA}%
  \BibitemOpen
  \bibfield  {author} {\bibinfo {author} {\bibfnamefont {K.}~\bibnamefont
  {Momma}}\ and\ \bibinfo {author} {\bibfnamefont {F.}~\bibnamefont {Izumi}},\
  }\bibfield  {title} {\bibinfo {title} {{VESTA 3 for three-dimensional
  visualization of crystal, volumetric and morphology data}},\ }\href
  {https://doi.org/10.1107/S0021889811038970} {\bibfield  {journal} {\bibinfo
  {journal} {J. Appl. Crystallogr.}\ }\textbf {\bibinfo {volume} {44}},\
  \bibinfo {pages} {1272} (\bibinfo {year} {2011})}\BibitemShut {NoStop}%
\bibitem [{\citenamefont {Okamoto}(1997)}]{Okamoto1997Ca-Sb}%
  \BibitemOpen
  \bibfield  {author} {\bibinfo {author} {\bibfnamefont {H.}~\bibnamefont
  {Okamoto}},\ }\bibfield  {title} {\bibinfo {title} {{Ca-Sb
  (Calcium-Antimony)}},\ }\href {https://doi.org/10.1007/BF02647863} {\bibfield
   {journal} {\bibinfo  {journal} {J. Phase Equilibria}\ }\textbf {\bibinfo
  {volume} {18}},\ \bibinfo {pages} {313} (\bibinfo {year} {1997})}\BibitemShut
  {NoStop}%
\bibitem [{\citenamefont {Yonezawa}\ \emph {et~al.}(2015)\citenamefont
  {Yonezawa}, \citenamefont {Higuchi}, \citenamefont {Sugimoto}, \citenamefont
  {Sow},\ and\ \citenamefont {Maeno}}]{Yonezawa2015ADR}%
  \BibitemOpen
  \bibfield  {author} {\bibinfo {author} {\bibfnamefont {S.}~\bibnamefont
  {Yonezawa}}, \bibinfo {author} {\bibfnamefont {T.}~\bibnamefont {Higuchi}},
  \bibinfo {author} {\bibfnamefont {Y.}~\bibnamefont {Sugimoto}}, \bibinfo
  {author} {\bibfnamefont {C.}~\bibnamefont {Sow}},\ and\ \bibinfo {author}
  {\bibfnamefont {Y.}~\bibnamefont {Maeno}},\ }\bibfield  {title} {\bibinfo
  {title} {{Compact AC susceptometer for fast sample characterization down to
  0.1 K}},\ }\href {https://doi.org/10.1063/1.4929871} {\bibfield  {journal}
  {\bibinfo  {journal} {Rev. Sci. Instrum.}\ }\textbf {\bibinfo {volume}
  {86}},\ \bibinfo {pages} {093903} (\bibinfo {year} {2015})}\BibitemShut
  {NoStop}%
\bibitem [{\citenamefont {Aharoni}(1998)}]{Aharoni1998DemagFactor}%
  \BibitemOpen
  \bibfield  {author} {\bibinfo {author} {\bibfnamefont {A.}~\bibnamefont
  {Aharoni}},\ }\bibfield  {title} {\bibinfo {title} {{Demagnetizing factors
  for rectangular ferromagnetic prisms}},\ }\href
  {https://doi.org/10.1063/1.367113} {\bibfield  {journal} {\bibinfo  {journal}
  {J. Appl. Phys.}\ }\textbf {\bibinfo {volume} {83}},\ \bibinfo {pages} {3432}
  (\bibinfo {year} {1998})}\BibitemShut {NoStop}%
\bibitem [{\citenamefont {Yang}\ \emph {et~al.}(1994)\citenamefont {Yang},
  \citenamefont {Kao}, \citenamefont {Xin},\ and\ \citenamefont
  {Wong}}]{Yang1994GrainBoundary}%
  \BibitemOpen
  \bibfield  {author} {\bibinfo {author} {\bibfnamefont {M.}~\bibnamefont
  {Yang}}, \bibinfo {author} {\bibfnamefont {Y.~H.}\ \bibnamefont {Kao}},
  \bibinfo {author} {\bibfnamefont {Y.}~\bibnamefont {Xin}},\ and\ \bibinfo
  {author} {\bibfnamefont {K.~W.}\ \bibnamefont {Wong}},\ }\bibfield  {title}
  {\bibinfo {title} {{Chemical doping and intergranular magnetic-field effects
  in bulk thallium-based superconductors}},\ }\href
  {https://doi.org/10.1103/PhysRevB.50.13653} {\bibfield  {journal} {\bibinfo
  {journal} {Phys. Rev. B}\ }\textbf {\bibinfo {volume} {50}},\ \bibinfo
  {pages} {13653} (\bibinfo {year} {1994})}\BibitemShut {NoStop}%
\bibitem [{\citenamefont {Kittaka}\ \emph {et~al.}(2009)\citenamefont
  {Kittaka}, \citenamefont {Nakamura}, \citenamefont {Yaguchi}, \citenamefont
  {Yonezawa},\ and\ \citenamefont {Maeno}}]{Kittaka2009Sr2RuO4}%
  \BibitemOpen
  \bibfield  {author} {\bibinfo {author} {\bibfnamefont {S.}~\bibnamefont
  {Kittaka}}, \bibinfo {author} {\bibfnamefont {T.}~\bibnamefont {Nakamura}},
  \bibinfo {author} {\bibfnamefont {H.}~\bibnamefont {Yaguchi}}, \bibinfo
  {author} {\bibfnamefont {S.}~\bibnamefont {Yonezawa}},\ and\ \bibinfo
  {author} {\bibfnamefont {Y.}~\bibnamefont {Maeno}},\ }\bibfield  {title}
  {\bibinfo {title} {{Spatial Development of Superconductivity in the
  \ce{Sr2RuO4}-Ru Eutectic System}},\ }\href
  {https://doi.org/10.1143/JPSJ.78.064703} {\bibfield  {journal} {\bibinfo
  {journal} {J. Phys. Soc. Jpn.}\ }\textbf {\bibinfo {volume} {78}},\ \bibinfo
  {pages} {064703} (\bibinfo {year} {2009})}\BibitemShut {NoStop}%
\bibitem [{\citenamefont {Tinkham}(2004)}]{Tinkham2004Superconductivity}%
  \BibitemOpen
  \bibfield  {author} {\bibinfo {author} {\bibfnamefont {M.}~\bibnamefont
  {Tinkham}},\ }\href@noop {} {\emph {\bibinfo {title} {{Introduction to
  Superconductivity}}}}\ (\bibinfo  {publisher} {Dover Publications},\ \bibinfo
  {address} {New York},\ \bibinfo {year} {2004})\BibitemShut {NoStop}%
\bibitem [{\citenamefont {Kobayashi}\ \emph {et~al.}(2016)\citenamefont
  {Kobayashi}, \citenamefont {Yanase},\ and\ \citenamefont
  {Sato}}]{Kobayashi2016NonsymmorphicSC}%
  \BibitemOpen
  \bibfield  {author} {\bibinfo {author} {\bibfnamefont {S.}~\bibnamefont
  {Kobayashi}}, \bibinfo {author} {\bibfnamefont {Y.}~\bibnamefont {Yanase}},\
  and\ \bibinfo {author} {\bibfnamefont {M.}~\bibnamefont {Sato}},\ }\bibfield
  {title} {\bibinfo {title} {{Topologically stable gapless phases in
  nonsymmorphic superconductors}},\ }\href
  {https://doi.org/10.1103/PhysRevB.94.134512} {\bibfield  {journal} {\bibinfo
  {journal} {Phys. Rev. B}\ }\textbf {\bibinfo {volume} {94}},\ \bibinfo
  {pages} {134512} (\bibinfo {year} {2016})}\BibitemShut {NoStop}%
\end{thebibliography}

%
\end{document}